\documentclass[plb,preprintnumbers,nofootinbib]{elsarticle}
\usepackage{bm}
\usepackage{latexsym}
\usepackage{dcolumn}
\usepackage{amsmath,amsfonts,amssymb}
\usepackage{graphicx,epsfig}
\usepackage{color}
\usepackage{amsthm}

\usepackage{lineno,hyperref}

\def\be {\begin{equation}}
\def\ee {\end{equation}}
\def\bea {\begin{eqnarray}}
\def\eea {\end{eqnarray}}
\def\bc {\begin{center}}
\def\ec {\end{center}}
\def\bfg {\begin{figure}}
\def\efg {\end{figure}}
\def\bi {\begin{itemize}}
\def\ei {\end{itemize}}

\def\a  {\alpha}
\def\b  {\beta}

\def\m  {\mu}
\def\n  {\nu}

\def\r  {\rho}

\def\beq{\begin{equation}}
\def\eeq{\end{equation}}
\def\br{\begin{eqnarray}}
\def\er{\end{eqnarray}}
\newcommand{\eel}[1] {\label{#1}\end{equation}}

\newcommand{\bdm}{\begin{displaymath}}
\newcommand{\edm}{\end{displaymath}}

\newcommand{\ket}[1]{|#1\rangle}

\begin{document}                             %updated 12 June 2011%

\begin{frontmatter}

\title{A gauge field theory of
fermionic\\ Continuous-Spin Particles}

%% Group authors per affiliation:
\author[Tours,Korea]{X. Bekaert}
\ead{xavier.bekaert@lmpt.univ-tours.fr}
\cortext[mycorrespondingauthor]{Corresponding author: X. Bekaert}

\author[Tours,Kurdistan]{M. Najafizadeh}
\ead{mnajafizadeh@gmail.com}

\author[Kurdistan]{M. R. Setare}
\ead{rezakord@ipm.ir}

\address[Tours]{Laboratoire de Math\'ematiques et Physique Th\'eorique\\
Unit\'e Mixte de Recherche $7350$ du CNRS\\
F\'ed\'eration de Recherche $2964$ Denis Poisson\\
Universit\'e Fran\c{c}ois Rabelais, Parc de Grandmont\\
37200 Tours, France}

\address[Korea]{B.W. Lee Center for Fields, Gravity and Strings\\ Institute for Basic Science\\ Daejeon, Korea}

\address[Kurdistan]{Department of Physics, Faculty of Sciences \\ University of Kurdistan \\66177-15177 Sanandaj, Iran}

\begin{abstract}
In this letter, we suggest a
local covariant action for a gauge field theory of fermionic Continuous-Spin Particles (CSPs).
The action is invariant under gauge transformations without any constraint on both the gauge field and the gauge transformation parameter.
The Fang-Fronsdal equations for a tower of massless fields with all half-integer spins
arise as a particular limit of the equation of motion of fermionic CSPs.
\end{abstract}

\begin{keyword}
Continuous Spin Particle, Poincar\'e Group Representation, Higher Spin Theory
\\
\texttt{arXiv:1506.00973 [hep-th]}
\end{keyword}

\end{frontmatter}

%\linenumbers

\section{Introduction}
    The unitary representations of the Poincar\'e group in four spacetime dimensions were first examined
    by E. Wigner in \cite{Wigner}.
    For massless particles, there is a class of representations, the so-called ``continuous-spin'' particles, %(CSPs),
    for which the eigenstates of different helicities are mixed under Lorentz transformations,
    similarly to the class of massive particles.
    In 3+1 dimensions, there exists only two types of CSP: the bosonic case where the spectrum of eigenvalues of the
    helicity operator is all the integers, and the fermionic case where the spectrum span all the half-integers.
    The helicity is defined, more covariantly, as
    $ W^2 \ket{%\hat
    {h}}= - \rho^2 \ket{%\hat
    {h}} $
    where $%\hat
    {h}$ is the helicity, %operator,
    $W^\m$ is the Pauli-Lubanski vector %operator
    and the real parameter $\rho$ (with the dimension of a mass) determines the degree of mixing of eigenstates.
    The eigenstates can be labeled by either integer or half-integer eigenvalues %of
    $%\hat
    {h}$, depending on the representation type.
    In the $\rho\to 0$ limit, the helicity-eigenstates reduce to the familiar ones that are Lorentz invariant, in the sense that they do not mix under Lorentz boosts (see e.g. \cite{ST 1} for more review).
    Recently, it was argued %illustrated
    that CSPs might evade the Weinberg no-go theorem on covariant soft emission amplitudes and
    could thus mediate long-range interactions \cite{ST 1}.

    As first pointed out by A.M. Khan and P. Ramond in \cite{KR}, one suggestive way to think about a CSP is as the limit of a massive particle where its mass $m$ goes to zero while its spin $s$ goes to infinity
       with their product being fixed ($m\to0$ and $s\to\infty$\,, with $ms=\rho$). This group-theoretical observation was translated at the field-theoretical level
       in \cite{BM} where Fronsdal-like equations of motion for bosonic CSPs and Fang-Fronsdal-like ones for fermionic CSPs
       where obtained from the above limit of the corresponding equations \cite{F 1,F 2} for massive higher-spin particles (see e.g. \cite{S} for a review) and shown to be equivalent to Wigner's equations \cite{wi} (see also \cite{Bargmann:1948ck} for more details).

 More recently, P. Schuster and N. Toro presented a local covariant action for
       bosonic CSPs, formulated with the help of an auxiliary Lorentz vector $\eta^\mu$ localized
       to the unit hyperboloid $\eta^2=-1$ \cite{ST 4}. This localization on a hyperboloid improved their initial proposal \cite{ST 3}
       and allows to recover precisely the equations of \cite{BM} as Euler-Lagrange equations.
See also the recent analysis of V. O. Rivelles \cite{R}.

       Until now, the gauge field theory of fermionic CSPs was missing from the literature at the level of the action.
       To describe supersymmetric CSP multiplets \cite{BR} or cross-interactions between bosonic and fermionic CSPs, it is unavoidable
       to construct an action of this type. The layout of the letter is as follows. In section \ref{II},
       a local covariant action of fermionic CSPs is proposed and we elaborate on its different aspects.
       In section \ref{III}, taking $\r=0$, Fang-Fronsdal equations %and results of higher-spin gauge fields
       for half-integer helicities will be obtained \cite{F 2}.
       We conclude and present open problems in section \ref{IV}.

       We will work in the ``mostly minus'' signature and focus on spacetime dimension four  but the higher ($D\geqslant 4$) and lower\footnote{There is a version in three spacetime dimensions of
       CSPs which can be thought as a massless generalization of anyons \cite{ST 5}.} dimensional generalizations are straightforward.

\section{Local and covariant action} \label{II}

We propose an action for the free fermionic CSPs as
\bea
S_{free} & = &   {\int  d^4 x \, d^4 \eta }  \,\,
 \Big[    \delta'(\eta^2 +1)\, \overline{\Psi}\, ( \gamma  \cdot {\eta} -  i ) (\gamma \cdot   \partial_x) \Psi  \nonumber \\
& & \qquad\quad\quad\quad\quad +\, \delta(\eta^2 +1) \overline{\Psi}\, \Delta  \Psi \Big]   \,, \label{Action}
\eea
where $\gamma^{\mu}$ are gamma matrices,
$\delta'(a)=\frac{d}{d a}\delta(a)$ and $\Delta = \partial_\eta\cdot\partial_x + \r$\,.

The gauge field $\Psi(\eta,x)$ is a spinor field, of which the spinor index has been omitted.
It is assumed that $\Psi$ is analytic in $\eta^\m$.
From the action, it is clear that $\Psi(\eta,x)$ has mass dimension $3/2$, as it should.
When $\r=0$, the helicity eigenstates factorize into
a tower of states with half-integer eigenvalues.
%
%In this sense, for $\r=0$ the helicity is invariant under Lorentz
%transformations while for $\r \neq 0$ the eigenstates mix under Lorentz
%boosts and (\ref{Action}) describe a single CSP.
%
%
The action is written in an enlarged spacetime where
inhomogeneous Lorentz transformations act on $x^{\mu}$ ($x'=\Lambda x +a$)
and homogeneous Lorentz transformations act on an auxiliary 4-vector coordinate $\eta^{\mu}$ ($\eta'=\Lambda \eta$).
%
%The spacetime metric is mostly minus.
%
The delta functions in (\ref{Action})
illustrate that the $\eta$ dependence of $\Psi(\eta,x)$ is localized to a unit hyperboloid in $\eta$-space,
an internal space that encodes spin. Note that no dynamics is carried out in $\eta$-space.
The action is invariant under the gauge transformation
\bea
\delta \Psi(\eta,x) &=& \Big[ (\gamma  \cdot \partial_x) (\gamma \cdot \eta + i)  - (\eta^2 +1) \Delta \Big]  \boldsymbol{\epsilon}(\eta,x)\nonumber\\
&&\quad+ \, (\eta^2 +1)( \gamma  \cdot {\eta} -  i )\boldsymbol{\chi}(\eta,x), \label{GT}
\eea
where $\boldsymbol{\epsilon}(\eta,x)$ and $\boldsymbol{\chi}(\eta,x)$ are arbitrary spinor gauge transformation parameters and there is no constraint on them.
The $\chi$ symmetry is the analogue of the one in \cite{ST 4, R} which allows us to remove the triple gamma-trace part of the gauge field.

In the presence of background currents, linear interactions can be given by
\bea
S_{int} & = & -i{ \int  d^4 x\, d^4\eta }\,\,
\delta'(\eta^2+1) \Big[\,\overline{\Psi} (\eta,x) (\gamma \cdot \eta - i) \boldsymbol{\sigma}(\eta,x) \nonumber \\
& & \quad\quad\quad\quad\quad-\,
\overline{{\boldsymbol{\sigma}}}(\eta,x) (\gamma \cdot \eta + i)  \Psi(\eta,x)\Big], \label{Intraction}
\eea
where $\boldsymbol{\sigma}$ and $\overline{\boldsymbol{\sigma}}$ \, are spinor sources.

The gauge invariance of $S_{int}$ leads to two continuity-like condition
\bea
\Big[\delta(\eta^2+1) (\gamma \cdot \eta - i) \,{\Delta} \Big]\boldsymbol{\sigma} (\eta,x)&=0,& \label{continuity condition}\\
\overline{\boldsymbol{\sigma}} (\eta,x) \Big[ \,\overleftarrow{\Delta}\, \delta(\eta^2+1) (\gamma \cdot \eta + i) \Big]&=0,&
\eea
for each source, where $\overleftarrow{\Delta}$ means that $\Delta$ operates to the left.
Using (\ref{Action}) and (\ref{Intraction}), it is straightforward to obtain a covariant equation of motion for the field $\Psi$
\bea
&\,\left[{\delta'(\eta^2+1)} (\gamma \cdot \eta - i ) (\gamma  \cdot \partial_x)
+ \, \delta(\eta^2+1) \, \Delta \right]\Psi&  \label{EOM}\\
&\qquad\qquad =i\,
{\delta'(\eta^2+1)} (\gamma \cdot \eta - i ) \boldsymbol{\sigma}. &\nonumber
\eea
As will be shown in the next section, this equation of motion describes a single fermionic CSP. In this approach, there is no constraint on the gauge field, contrarily to the Fang-Fronsdal formulation (see \cite{FS} for a local unconstrained formulation of massless higher-spin fields).

One of the main purposes of this paper is to present a local and covariant action for the fermionic CSPs which reproduces fermionic higher-spin massless particles in the $\r\to0$ limit (called ``helicity correspondence'' in \cite{ST 4}).
To demonstrate this connection, we shall transform our equations to those in $\omega$-space, the conjugate space of the $\eta$-space. In $\omega$-space, we will show that our equation of motion is %very similar
equivalent to the Fang-Fronsdal-like equation \cite{BM}, which was obtained from the massive Fang-Fronsdal equation and is equivalent to the Wigner equations \cite{wi}.
\section{Relation to the Fang-Fronsdal equation} \label{III}
We perform a Fourier transformation in $\eta^\m $ to express the Grassmann
variables in the $\omega$-space as
\bea
\Psi(\omega,x)&\equiv& \int d^4\eta \,e^{i\eta\cdot\omega}
 \delta'(\eta^2 +1) (\gamma \cdot \eta + i)\Psi(\eta,x),  \label{1} \\
%\ee
%\be
\boldsymbol{\sigma}(\omega,x)&\equiv& \int d^4\eta \,e^{i\eta\cdot\omega}
 \delta'(\eta^2 +1) (\gamma \cdot \eta - i)\boldsymbol{\sigma}(\eta,x),\\
%\ee
%\be
\boldsymbol{\epsilon}(\omega,x)&\equiv& \int d^4\eta \, e^{i\eta\cdot\omega}
 \delta(\eta^2 +1) (\gamma \cdot \eta + i)\boldsymbol{\epsilon}(\eta,x).  \label{2}
\eea
Notice that the fields in the left-hand-sides are unconstrained while the ones in the right-hand-side are constrained.
%and similarly for $\boldsymbol{\sigma}$ and $\boldsymbol{\epsilon}$.
%
%Considering the above Fourier transformation, we can find useful transformations for $\Psi$, $\boldsymbol{\sigma}$ and $\boldsymbol{\epsilon}$. For instance, we can obtain the following transformations for the gauge field $\Psi$,
%\be
%\delta'(\eta^2 +1) (\gamma \cdot \eta + i)\Psi(\eta,x)
%=
%\int d^4\omega \,e^{-i\eta\cdot\omega}\,\Psi(\omega,x), \label{1}
%\ee
%\be
%\delta(\eta^2 +1) (\gamma \cdot \eta + i)\Psi(\eta,x)
%=
%\int d^4\omega \,e^{-i\eta\cdot\omega} (  \partial^2 _{{\omega}} -1)\,\Psi(\omega,x), \label{2}
%\ee
%\be
%\delta(\eta^2 +1) \Psi(\eta,x)
%=
%i\,\int d^4\omega \,e^{-i\eta\cdot\omega} (\c^\m \partial_{{\omega}_\m} +1)\,\Psi(\omega,x),\label{3}
%\ee
%
More precisely, the equations \eqref{1} and \eqref{2} can be understood as the general solutions of the triple gamma-trace condition
 \be
 \left(\gamma\cdot \partial_{\omega} + 1\right)\left(
\partial_{\omega}\cdot \partial_{\omega} - 1\right)\,\Psi(\omega,x)=0,
 \ee
 and the gamma-trace condition
 \be
 \left(\gamma \cdot \partial_{\omega} + 1 \right)\boldsymbol{\epsilon}(\omega,x)=0,
 \ee
 which are equivalent to the ones in \cite{BM} (up to a multiplication by the matrix $i\gamma^5$ as explained below).
Let us point out that the fields in the left-hand-sides of \eqref{1} and \eqref{2} do not uniquely determine
the fields $\Psi(\eta,x)$ and $\boldsymbol{\epsilon}(\eta,x)$ in the right-hand-side, but only up to some gamma-trace terms. In particular, the arbitrariness in the  field $\Psi(\eta,x)$
is nothing but the $\chi$ symmetry in \eqref{GT}. In other words, the field $\Psi(\omega,x)$ is not affected by the $\chi$ symmetry.
As one can check, the change of variables \eqref{1}-\eqref{2} converts some of the gauge symmetries of the original fields (e.g. the $\chi$ symmetry) into 
conditions imposed on the new fields (e.g. gamma-trace constraint). This fact is closely related to the standard conversion of first-class constraints into second-class ones.

Multiplying the equation (\ref{GT}) by $\delta'(\eta^2 +1) (\gamma \cdot \eta +i)$ to the left, we obtain
\be
 \delta'(\eta^2 +1) (\gamma \cdot \eta +i) \delta \Psi(\eta ,x) = \Delta \big[\delta(\eta^2 +1) (\gamma \cdot \eta +i)  \boldsymbol{\epsilon}(\eta,x) \big].
 \label{intermediate}
\ee
Now, Fourier transforming \eqref{intermediate} over the auxilliary variable $\eta$,
the gauge transformation takes the form of
\be
\delta\Psi(\omega,x)=
({\omega \cdot \partial_x} + i\r)   \boldsymbol{\epsilon}(\omega,x), \label{GT omega}
\ee
where a constant factor has been absorbed in the gauge field. This is exactly the gauge transformation of the Fang-Fronsdal-like equation, proposed in \cite{BM}.
Let us stress that the $\chi$ symmetry is absent in \eqref{GT omega}.

The continuity condition (\ref{continuity condition}) appears in $\omega$-space  as
\be
\Big[ (\gamma \cdot \partial_{\omega} +1 ) (\gamma \cdot \partial_{x}) - ({\omega \cdot \partial_x} + i\r) \big(\partial_{\omega}^2 -1 \big)
 \Big] \boldsymbol{\sigma}(\omega,x)=0. \label{continuity condition omega}
\ee
The equation of motion (\ref{EOM}) turns into
\be
i\Big[ (\gamma\cdot\partial_x) %{ \gamma^{\mu}} { \partial_{\mu} }
-
(\omega \cdot \partial_x + i\r) (\gamma\cdot {\partial_{\omega}} +1 )
  \Big]\Psi(\omega,x)= \boldsymbol{\sigma}(\omega,x). \label{EOM omega}
\ee
A gauge invariant equation of motion equivalent to
(\ref{EOM omega}),
%, up to a multiplication by the matrix $i\gamma^5$ (with $\boldsymbol{\sigma}=0$),
with $\boldsymbol{\sigma}=0$, was obtained in \cite{BM} from the massless high-spin limit of the equation for fermionic massive particles,
but no action leading to this equation of motion was presented. To see the equivalence between \eqref{EOM omega} and the equation written in \cite{BM}, we can multiply \eqref{EOM omega} by the matrix $i\gamma^5$ to the left
 (with $\boldsymbol{\sigma}=0$) and get
 \be
 \Big[ (\Gamma\cdot\partial_x) %{ \gamma^{\mu}} { \partial_{\mu} }
-
(\omega \cdot \partial_x + i\r) (\Gamma\cdot {\partial_{\omega}} + i\,\Gamma^5  )
  \Big]\Psi(\omega,x)= 0,  \label{22}
 \ee
 where $\Gamma^\m = i \gamma^5 \gamma ^\m$ and $\Gamma^5=\gamma^5$. These new matrices $\Gamma$'s satisfy the same Clifford algebra as the original matrices $\gamma$'s and the obtained equation is the one in \cite{BM}.\footnote{Notice that the mostly plus signature was used in \cite{BM} and is responsible for a distinct $i$ factor.}
%responsible for a relative factor $i$

 Via a gauge-fixing procedure similar to the one in \cite{BM}, one can show that the equation \eqref{EOM omega} without source describes a single fermionic CSP.
In fact, we can impose the gauge
\be
(\gamma\cdot {\partial_{\omega}} +1 )
\Psi(\omega,x)= 0, \label{gfix}
\ee
and get from \eqref{EOM omega} with $\boldsymbol{\sigma}=0$:
\be
i (\gamma\cdot\partial_x){\Psi}(\omega,x)=0. \label{EOMgfix}
\ee
In turn, the equations \eqref{gfix} and \eqref{EOMgfix} imply that
\be
(\partial_\omega\cdot\partial_x)\Psi(\omega,x)= 0\,.\label{compatibility}
\ee
As explained in \cite{BM}, these three equations \eqref{gfix}-\eqref{compatibility} are equivalent to Wigner's equations \cite{wi} which are known to describe a single fermionic CSP.

To make contact between the above equations and the corresponding Fang-Fronsdal equations, one can first rescale\footnote{X.~B. is grateful to J. Mourad for discussions on the corresponding rescaling in the bosonic case.}  the auxilliary variable (and the gauge parameter) as follows:
$\omega\to \rho^\frac12\omega$ in (\ref{GT omega})-(\ref{EOM omega}) and then put $\r=0$.
For instance, (\ref{EOM omega}) reads in terms of the rescaled variable as
\be
i\Big[ (\gamma\cdot\partial_x) %{ \gamma^{\mu}} { \partial_{\mu} }
-
(\omega \cdot \partial_x + i\r^\frac12) (\gamma\cdot {\partial_{\omega}} +\r^\frac12 )
  \Big]\Psi(\omega,x)= \boldsymbol{\sigma}(\omega,x),
\ee
which in the $\rho\to0$ limit leads to the Fang-Fronsdal equation
\be
i\Big[ (\gamma\cdot\partial_x) %{ \gamma^{\mu}} { \partial_{\mu} }
-
(\omega \cdot \partial_x ) (\gamma\cdot {\partial_{\omega}})
  \Big]\Psi(\omega,x)= \boldsymbol{\sigma}(\omega,x), \label{EOM omega'}
\ee%
Similarly, one gets from \eqref{continuity condition omega} in the same $\rho\to0$ limit
\be
\Big[ (\gamma \cdot \partial_{\omega}) (\gamma \cdot \partial_{x}) - ({\omega \cdot \partial_x}) %\big(
\partial_{\omega}^2 %-1 \big)
 \Big] \boldsymbol{\sigma}(\omega,x)=0. \label{continuity condition omega'}
\ee
The spinor field $\Psi$
%$\boldsymbol{\sigma}$
%and $\boldsymbol{\epsilon}$)
can be considered of the form
\be
\Psi(\omega,x) = \psi(x)
+
\omega^{\mu}\psi_{\mu}(x)
+
\frac{1}{2}\,\omega^{\mu}\omega^{\nu}\psi_{\mu\nu}(x)
+
\cdots,
%\frac{1}{3!}\omega^{\mu}\omega^{\nu}\omega^{\rho}\theta_{\mu\nu\rho}+\cdots,
\ee
where
$\psi$ is a spinor (Dirac) field of helicity $\tfrac{1}{2}$,
$\psi_{\mu}$ is a vector-spinor (Rarita-–Schwinger) field of helicity $\tfrac{3}{2}$,
$\psi_{\mu\nu}$ is a symmetric tensor-spinor field of helicity $\tfrac{5}{2}$, etc.
We will have the same definition for the spinor field $\boldsymbol{\sigma}$ as above.
For $\boldsymbol{\epsilon}$ we can write
\be
\boldsymbol{\epsilon}(\omega,x) = \epsilon(x)
+
\omega^{\mu}\epsilon_{\mu}(x)
+
\frac{1}{2}\,\omega^{\mu}\omega^{\nu}\epsilon_{\mu\nu}(x)
+
\cdots,
%\frac{1}{3!}\omega^{\mu}\omega^{\nu}\omega^{\rho}\theta_{\mu\nu\rho}+\cdots,
\ee
where $\epsilon$ is the gauge parameter of the helicity $\tfrac{3}{2}$ gauge field and so on.
By assuming the fields analytic in $\omega$-space, the Fang-Fronsdal formulation of half-integer spin gauge fields can conveniently be elaborated as follows:

 According to (\ref{GT omega}) at $\rho=0$, one can see that the Dirac field is not a gauge field, but all other
 massless fields transform under the gauge symmetries.
 The gauge transformations for $ s = \tfrac{3}{2} , \tfrac{5}{2},\cdots$, take
 the standard form \cite{F 2}
 \bea
 %(\omega ^\m )
 \delta \psi_{\m} & = & %(\omega ^\m)
 \partial_{\m} \epsilon, \nonumber \\
% \big(\tfrac{1}{2}\omega^{\mu}\omega^{\nu} \big)
\delta \psi_{\m\n}
  & = &
 %\big( \tfrac{1}{2}\omega^{\mu}\omega^{\nu}\big)
 %(
 \partial_{\m}\epsilon_{\n}+\partial_{\n}\epsilon_{\m}%)
 ,\\
  &\vdots&\nonumber
  \eea
The equation of motion \eqref{EOM omega'}, for $ s= \tfrac{1}{2}, \tfrac{3}{2}, \tfrac{5}{2}, \cdots $ reduce to
\bea
i\,({\gamma\cdot\partial}) \psi  & = & \sigma, \nonumber \\
%
%\omega^\a
i\,\Big(
({\gamma\cdot\partial}) \psi_{\alpha}
-\partial_{\alpha} \psi^\prime\Big)
& = & \sigma_{\a}, \nonumber \\
%
%\tfrac{1}{2} \omega^\a \omega^\b
i\,\Big(
({\gamma\cdot\partial}) \psi_{\alpha \beta}
-\partial_{\alpha}%\gamma^{\mu}
\psi^\prime_{%\mu
\beta}
-\partial_{\beta}%\gamma^{\mu}
\psi^\prime_{ \alpha %\mu
}\Big)
& = & \sigma_{\a\b}, \nonumber \\
&  \vdots & \nonumber
\eea
which are exactly Fang-Fronsdal equations for half-integer higher-spin gauge fields \cite{F 2}.
The Fang-Fronsdal notation for trace has been used ($\gamma$-trace %$\epsilon\equiv\epsilon ',
$\epsilon '_{\nu\rho\,\cdots} \equiv \gamma^{\m} \epsilon_{\mu\nu\rho\,\cdots} $).
%
%
% In addition,

Ultimately, the %familiar
continuity conditions (for $ s=\tfrac{3}{2}, \tfrac{5}{2}, \cdots$) can be extracted
from (\ref{continuity condition omega'})
\bea
\partial^\m \sigma_{\m}&=&\tfrac12\gamma^\m\partial_\m\sigma^\prime, \nonumber\\
\partial^\n \sigma_{\m\n}&=&\tfrac{1}{2}\big(\gamma^\n\partial_\n \sigma^\prime_{\m}+\partial_\m\sigma^{\prime\prime}\big),
 \\
&  \vdots & \nonumber
\eea
which indeed correspond to the ones in \cite{F 2}.
\\
\section{Conclusions and Discussion} \label{IV}
%
%\vspace{7cm}
%
In this letter, we proposed a local, covariant and gauge-invariant, action \eqref{Action} to describe
fermionic CSPs.
As is standard in higher-spin litterature, an auxiliary Minkowski space ($\eta$-space here) was used to encode spinning degrees of freedom. However, there is no dynamics within the $\eta$-space.
We rewrote, in the conjugate $\omega$-space, the gauge symmetries and the equation of motion, and related them to the ones in \cite{BM}.
Finally, taking a suitable $\r\to 0$ limit, the Fang-Fronsdal equations for fermionic higher spin gauge fields were correctly obtained.

The fermionic CSP action proposed here, together with the bosonic CSP action of Schuster and Toro action, may open a new window to probe supersymmetric CSPs, or investigate Yukawa-like interactions of CSPs.
We let the canonical and path integral quantizations of fermionic CSPs for future work.

\section*{Acknowledgments}

We acknowledge helpful discussions with Dmitri Sorokin and Victor Rivelles. X.B. also thanks Jihad Mourad, Jeong-Hyuck Park and Denia Polydorou for useful exchanges.
M.N. thanks the University of Tours for hospitality. The research of X.B. was supported by the Russian Science Foundation grant 14-42-00047 in association with Lebedev Physical Institute.

% \newpage

%

\end{document}